\documentclass[runningheads]{llncs}
\usepackage[T1]{fontenc}
\usepackage{graphicx}
\usepackage{xcolor}
\usepackage{subfigure}

\usepackage{amsmath, amssymb, amsfonts, bm, mathrsfs, mathtools}

\usepackage{subcaption}  
\usepackage{comment}     
\usepackage{ulem}        
\usepackage{url}         
\usepackage{multirow}    

\usepackage{algorithm}
\usepackage{algorithmicx}
\usepackage{algpseudocodex}
\usepackage{booktabs}
\usepackage{subfigure}

\begin{document}
\title{Subset Selection for Stratified Sampling \\ 
in Online Controlled Experiments}
%
%
\author{Haru Momozu\inst{1}\orcidID{0009-0003-7450-7031}\and
Yuki Uehara\inst{2}\orcidID{0009-0001-8940-8461}\and
Naoki Nishimura\inst{1,3}\orcidID{0000-0002-6906-4323}\and
Koya Ohashi\inst{4}\orcidID{0009-0005-3356-3613}\and 
Deddy Jobson\inst{4}\orcidID{0000-0003-1557-8131}\and
Yilin Li\inst{4}\orcidID{0009-0002-3765-0755}\and
Phuong Dinh\inst{4}\orcidID{0009-0006-8682-1070}\and
Noriyoshi Sukegawa\inst{5}\orcidID{0000-0002-3560-0036}\and
Yuichi Takano\inst{1}\orcidID{0000-0002-8919-1282}}
\authorrunning{H. Momozu et al.}
%
\institute{University of Tsukuba, Tsukuba, Ibaraki 305-8573 Japan\\
\email{s2110892@u.tsukuba.ac.jp, ytakano@sk.tsukuba.ac.jp}\and 
Preferred Networks, Inc., Chiyoda-ku, Tokyo 100-0004 Japan\\
\email{yukiuehara00@preferred.jp}\and
Recruit Co., Ltd, Chiyoda-ku, Tokyo 100-6640 Japan\\
\email{nishimura@r.recruit.co.jp}\and
Mercari, Inc., Minato-ku, Tokyo 106-6118 Japan\\
\email{s2540404@u.tsukuba.ac.jp, \{deddy,y-li,pdinh\}@mercari.com}\and
Hosei University, Koganei-shi, Tokyo 184-8584 Japan\\
\email{sukegawa@hosei.ac.jp}}
%
\maketitle              
\begin{abstract} 
Online controlled experiments, also known as A/B testing, are the digital equivalent of randomized controlled trials for estimating the impact of marketing campaigns on website visitors. 
Stratified sampling is a traditional technique for variance reduction to improve the sensitivity (or statistical power) of controlled experiments; this technique first divides the population into strata (homogeneous subgroups) based on stratification variables and then draws samples from each stratum to avoid sampling bias. 
To enhance the estimation accuracy of stratified sampling, we focus on the problem of selecting a subset of stratification variables that are effective in variance reduction. 
We design an efficient algorithm that selects stratification variables one by one by simulating a series of stratified sampling processes. 
We also estimate the computational complexity of our subset selection algorithm. 
Computational experiments using synthetic and real-world datasets demonstrate that our method can outperform other variance reduction techniques especially when multiple variables have a certain correlation with the outcome variable. 
Our subset selection method for stratified sampling can improve the sensitivity of online controlled experiments, thus enabling more reliable marketing decisions.

\end{abstract}
\section{Introduction}

\subsection{Background}

\begin{figure}[t]
\centering
\includegraphics[width=0.9\linewidth]{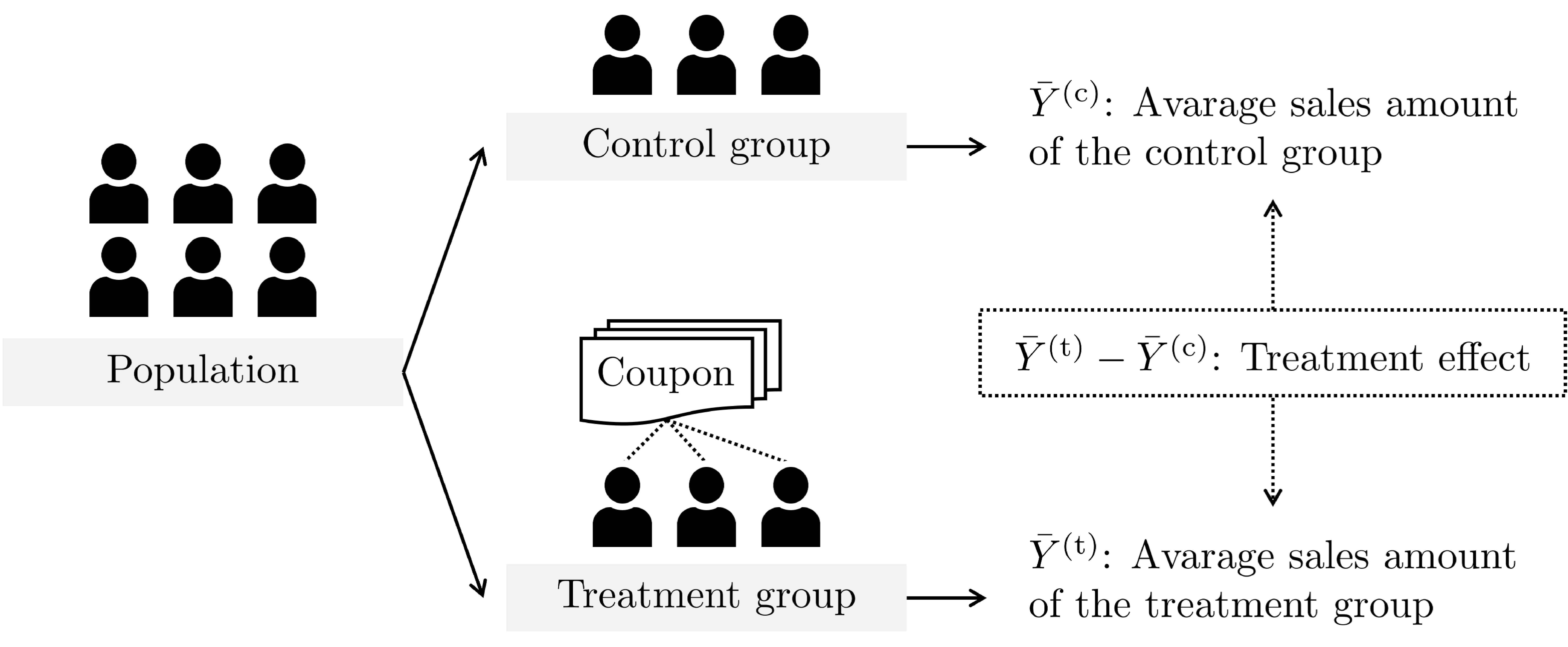}
\caption{Online controlled experiment for estimating coupon effects}
\label{fig:ex}
\end{figure}

A randomized controlled trial (RCT) is an experimental method for estimating the treatment effect by randomly dividing subjects into treatment and control groups and giving the treatment only to the treatment group.
RCTs have been considered the gold standard for providing evidence of causal relationships between treatments and outcomes~\cite{bhide2018simplified}.
Online controlled experiments (OCEs), also known as A/B testing, are the digital equivalent of RCTs to estimate the impact of marketing campaigns, such as a coupon distribution~\cite{ohashi2024strategic,uehara2024robust}, 
on website visitors~\cite{larsen2024statistical} (Fig.~\ref{fig:ex}). 
A distinctive feature of OCEs is that the collected user data can be utilized to design controlled experiments~\cite{CUPED}.
OCEs are widely practiced by major technology companies such as Google, Meta, LinkedIn, and Microsoft~\cite{quin2024b}. 

On websites with a large number of visitors, even small differences can have a significant impact on key metrics~\cite{larsen2024statistical}. 
Against this background, one of the important challenges of OCEs is to improve the sensitivity (or statistical power) of experiments, or in other words, to improve the ability of the experiment to detect treatment effects that actually exist.
The simplest way to improve the sensitivity is to increase the sample size of subjects included in the experiment. 
However, repeating experiments on a large number of visitors is likely to negatively impact the user experience on the website~\cite{COSS}.
It is therefore desirable to increase the sensitivity of experiments without increasing the sample size. 

\subsection{Related Work}
Variance reduction techniques have been used effectively to improve the accuracy of estimates obtained by Monte Carlo sampling~\cite{kroese2013handbook}.
Typical variance reduction techniques used to improve the sensitivity of controlled experiments can be categorized into two types~\cite{CUPED}: 
control variates and stratified sampling.

Methods of control variates reduce the variance of treatment effect estimates by expressing the outcome variable as a regression model of covariates~\cite{lin2013agnostic}.
This technique is also known as CUPED (controlled experiments using pre-experiment data)~\cite{CUPED}, which has become a standard tool in OCEs.
Guo et al.~\cite{MLRATE} proposed MLRATE (machine learning regression-adjusted treatment effect estimator), a control variates method that leverages cross-validated machine learning predictions. 
Jobson et al.~\cite{COSS} proposed COSS (covariate ordered systematic sampling), which alternately samples treatment and control groups according to the order of covariate values.

Stratified sampling is a traditional technique for variance reduction, which first divides the population into strata (homogeneous subgroups) based on stratification variables and then draws samples from each stratum to avoid sampling bias~\cite{stratifiedsampling}. 
Clustering techniques such as $K$-means clustering have been used for stratification~\cite{clustering_stratification,kim2013stratified}.
Several optimization algorithms have been developed to calculate the optimal sample size from each stratum~\cite{brito2024mathematical,optimal}.
Estimation of treatment effects using individual-level variance estimates was also considered~\cite{liou2020variance}.
The Netflix case study~\cite{netflix} demonstrated that three variance reduction methods (stratified sampling, post-stratification, and CUPED) contribute to improving the sensitivity of OCEs.

To the best of our knowledge, however, none of the prior studies have explored algorithms dedicated to selecting a subset of stratification variables that are effective in variance reduction.
Various methods have been proposed to select a subset of variables used for clustering~\cite{featureselection}. 
These subset selection methods help improve the accuracy, computational efficiency, interpretability, and robustness of clustering by identifying variables required for proper grouping.

\subsection{Contribution}

The motivation behind this research is to improve the variance reduction performance of stratified sampling by applying a subset selection algorithm to multivariate datasets.
For example, let us consider a coupon that young people respond strongly to. 
In this case, the coupon effect will be underestimated if a large number of elderly people are selected for the treatment group through simple random sampling. 
In addition to age, other variables such as gender, place of residence, and purchase history may also be correlated with the coupon effect, so it is crucial to appropriately select these stratification variables in stratified sampling.

A main goal of this paper is to establish a computational framework for selecting a subset of stratification variables for variance reduction. 
For this purpose, we design an efficient algorithm for subset selection based on the sequential forward search~\cite{wrapper}. 
Specifically, our method selects stratification variables one by one by simulating a series of stratified sampling processes. 
We also evaluate the computational complexity of our subset selection algorithm. 

To validate the effectiveness of our method, we conducted computational experiments using synthetic and real-world datasets. 
Experimental results demonstrate that our method can select stratification variables that are effective for variance reduction in stratified sampling. 
Moreover, our method can outperform other variance reduction methods especially when multiple variables have a certain correlation with the outcome variable. 

\section{Online Controlled Experiments}

Let $\bar{Y}^{\mathrm{(t)}}$ and $\bar{Y}^{\mathrm{(c)}}$ be the sample means of the outcome variable (e.g., sales amount, number of conversions, etc.) in the treatment and control groups, respectively. 
The treatment effect is then quantified by the average treatment effect:
\begin{equation}
\bar{Y}^{\mathrm{(t)}} - \bar{Y}^{\mathrm{(c)}}. 
\end{equation} 

The two-sample $t$-test is often conducted to test for significant differences between treatment and control groups.
With the null hypothesis $H_0: \bar{Y}^{\mathrm{(t)}}-\bar{Y}^{\mathrm{(c)}}=0$, the $t$-statistic is defined as
\begin{equation}\label{eq:t-stat}
t = \frac{\bar{Y}^{\mathrm{(t)}}-\bar{Y}^{\mathrm{(c)}}}{\sqrt{\mathrm{Var}(\bar{Y}^{\mathrm{(t)}}-\bar{Y}^{\mathrm{(c)}})}},
\end{equation}
where $\mathrm{Var}(\,\cdot\,)$ denotes the variance of an estimate resulting from random sampling.

To improve the sensitivity (or statistical power) of experiments, we need to increase the $t$-statistic in Eq.~\eqref{eq:t-stat} by decreasing the variance in the denominator.
Since the two samples are independent, the variance is rewritten as 
\begin{equation}
\mathrm{Var}(\bar{Y}^{\mathrm{(t)}}-\bar{Y}^{\mathrm{(c)}}) = \mathrm{Var}(\bar{Y}^{\mathrm{(t)}}) + \mathrm{Var}(\bar{Y}^{\mathrm{(c)}}).
\end{equation}
This implies that improving the sensitivity is equivalent to reducing the outcome variance for each group.

\section{Stratified Sampling}

In this section, we explain the three processes of stratified sampling: stratification, sample allocation, and calculation of the sample mean.
In what follows, we denote the set of consecutive integers as $[n] \coloneqq \{1,2,\ldots,n\}$.

\subsection{Stratification}
Let $N$ be the population size of subjects (e.g., all members of a website).
Stratification is a process of dividing the population into strata (homogeneous subgroups) based on stratification variables. 
Typically, a single covariate such as age, gender, or race is used for stratification~\cite{stratifiedsampling}. 
Clustering methods are also very effective when multiple variables are used for stratification~\cite{clustering_stratification,kim2013stratified}. 
As a result of stratification, the size $N_k$ of each stratum $k \in [K]$ is determined, such that $N = \sum_{k=1}^K N_k$. 

\subsection{Sample Allocation}
Let $n$ be the sample size, which is preferably much smaller than the population size $N$.  
Sample allocation is the process of allocating an appropriate sample size $n_k$ to each stratum $k \in [K]$, such that $n = \sum_{k=1}^K n_k$. 

Proportional sample allocation draws samples according to the proportion of each stratum~\cite{stratifiedsampling}. 
The sample size from each stratum is given by 
\begin{equation}\label{eq:prop_alloc}
n_k \approx \frac{N_k}{N} n \quad(k \in [K]).
\end{equation}

Optimal sample allocation determines the optimal sample sizes $\bm{n} \coloneqq (n_k)_{k \in [K]} \in \mathbb{Z}^K_{+}$ such that the variance of the sample mean is minimized. 
Specifically, it amounts to solving the following integer optimization problem~\cite{optimal}:  
\begin{equation}\label{eq:opt_alloc}
\min_{\bm{n} \in \mathbb{Z}_+^K} \quad \sum_{k=1}^K \frac{N_k^2 \sigma_k^2}{n_k} \quad 
\text{s.~t.} \quad \sum_{k=1}^K n_k = n, \quad 
\ell_k \leq n_k \leq u_k \quad (k \in [K]),
\end{equation}
where $\sigma_k^2$ is the outcome variance, and $\ell_k$ and $u_k$ are respectively the lower and upper bounds on the sample size for each stratum $k \in [K]$. 

\subsection{Calculation of the sample mean}

Let $\bar{Y}_k$ be the sample mean of the outcome variable for each stratum $k \in [K]$. 
The overall sample mean is then calculated by the weighted average: 
\begin{equation}\label{eq:sample_mean}
\hat{Y}_{\rm strat} = \sum_{k=1}^K \frac{N_k}{N} \bar{Y}_k.
\end{equation}

It is known (e.g., in Friedrich et al.~\cite{optimal}) that the variance of the sample mean in Eq.~\eqref{eq:sample_mean} is calculated with the finite population correction as 
\begin{equation}\label{eq:var}
\text{Var}(\hat{Y}_{\rm strat}) = \frac{1}{N^2} \left( \sum_{k=1}^{K} \frac{N_k^2 \sigma_k^2}{n_k} - \sum_{k=1}^{K} N_k \sigma_k^2 \right).
\end{equation}
It is also known (e.g., in Xie and Aurisset~\cite{netflix}) that the variance of the sample mean in stratified sampling is smaller than that in simple random sampling by $\sum_{k=1}^K N_k (\mu_k - \mu)^2/(nN)$, where $\mu$ is the population mean of the outcome variable, and $\mu_k$ is that in each stratum $k \in [K]$. 

\section{Subset Selection for Stratification}
In this section, we present our algorithm for selecting an effective subset of stratification variables. 
We also discuss the computational complexity of our algorithm.

\subsection{Sequential Forward Search for Variance Reduction}

We focus on the problem of selecting $\theta$ variables useful for stratification from $p$ candidate variables. 
To this end, we design an algorithm based on the sequential forward search~\cite{wrapper}, which selects variables one by one while evaluating its clustering performance.
Algorithm~\ref{alg1} summarizes our subset selection algorithm for stratified sampling.

Let $\mathcal{F} \subseteq [p]$ be an incumbent subset of stratification variables, and $V(\mathcal{F})$ be an evaluation metric defined by the variance in Eq.~\eqref{eq:var}. 
Our algorithm starts with the empty set $\mathcal{F} \leftarrow \emptyset$ and its variance $V(\mathcal{F}) \coloneqq +\infty$. 

Next, we repeat the following processes for each unselected variable $f \in [p]\setminus \mathcal{F} $:  
\begin{itemize}
    \item \textbf{Stratification}: Perform $K$-means clustering with the subset $\mathcal{F} \cup \{f\}$ of variables to divide the population into $K$ strata; 
    \item \textbf{Sample allocation}: Determine the sample sizes $n_k$ for $k \in [K]$ using the proportional allocation in Eq.~\eqref{eq:prop_alloc} or the optimal allocation in Eq.~\eqref{eq:opt_alloc}; 
    \item \textbf{Variance evaluation}: Calculate the variance in Eq.~\eqref{eq:var} to define $V(\mathcal{F} \cup \{f\})$. 
\end{itemize}

We then select one of the unselected variables, $f^\star \in [p] \setminus \mathcal{F} $, such that the corresponding variance $V(\mathcal{F} \cup \{f^\star\})$ is the smallest. 
If the variance is reduced, we update the incumbent subset as $\mathcal{F} \leftarrow \mathcal{F} \cup \{f^\star\}$ and return to the process of evaluating unselected variables. 
If the variance is not reduced, we terminate the algorithm with the incumbent subset $\mathcal{F} \subseteq [p]$.
We repeat these processes until the subset size is equal to $\theta$.

\begin{algorithm}[t]
\caption{Sequential Forward Search for Variance Reduction}
\label{alg1}
\textbf{Input:} 
Subset size $\theta \in \mathbb{N}$, number of strata $K \in \mathbb{N}$. \\
\textbf{Initialize:} Subset of variables $\mathcal{F} \leftarrow \emptyset$, evaluation metric $V(\mathcal{F}) \coloneqq +\infty$. 
\begin{algorithmic}[1]
\While{$|\mathcal{F}| < \theta$}
    \ForAll{$f \in [p] \setminus \mathcal{F} $} 
        \State Perform $K$-means clustering with $\mathcal{F} \cup \{f\}$. \Comment{stratification}
        \State Determine $n_k$ for $k \in [K]$. \Comment{sample allocation}
        \State Calculate $V(\mathcal{F} \cup \{f\})$ based on Eq.~\eqref{eq:var}. \Comment{variance evaluation}
    \EndFor
    \State Select $f^\star \in \arg \min \{V(\mathcal{F} \cup \{f\}) \mid f \in [p] \setminus \mathcal{F} \}$. 
    \If{$V(\mathcal{F} \cup \{f^\star\}) < V(\mathcal{F})$}
        \State Update $\mathcal{F} \leftarrow \mathcal{F} \cup \{f^\star\}$. 
    \Else
        \State \textbf{break}
    \EndIf
\EndWhile
\end{algorithmic}

\textbf{Output:} Subset of variables $\mathcal{F} \subseteq [p]$.
\end{algorithm}

\subsection{Computational Complexity}\label{sec:complexity}

A naive estimate of the computational complexity of $K$-means clustering is $\mathcal{O}(KNpT)$, where $K$ is the number of clusters, $N$ is the number of data points, $p$ is the number of variables, and $T$ is the number of iterations~\cite{kmeanstime}. 
As mentioned in Pakhira~\cite{pakhira2014linear}, this estimate can be rewritten as $\mathcal{O}(N^2 p)$ if we assume that $K$ is a constant and $T$ is proportional to $N$. 

The problem (Eq.~\eqref{eq:opt_alloc}) for optimal sample allocation can be solved in $\mathcal{O}(K \cdot \log_2 K \cdot \log_2 (n/K))$ time using the capacity scaling algorithm based on the polymatroidal structure of the feasible region~\cite{optimal}. 
Moreover, assuming that $K$ is a constant reduces the computational complexity to $\mathcal{O}(\log_2 n)$, which is smaller than $\mathcal{O}(N^2 p)$.

The sequential forward search calculates the evaluation metric $\mathcal{O}(p \theta)$ times~\cite{wrapper} and performs $K$-means clustering and sample allocation for each evaluation.
As a result, the computational complexity of Algorithm~\ref{alg1} is estimated to be $\mathcal{O}(N^2 p^2 \theta)$, or $\mathcal{O}(N^2 p^2)$ if we assume that $\theta$ is a constant.

\section{Experiments}

In this section, we report experimental results to evaluate the effectiveness of our subset selection method for stratified sampling. 

\subsection{Experimental Setup}
We compared the performance of the following methods for variance reduction:
\begin{itemize}
    \item \textbf{CUPED}: Control variates method using pre-experiment data~\cite{CUPED}; 
    \item \textbf{COSS}: Covariate ordered systematic sampling~\cite{COSS}; 
    \item \textbf{K-means}: Stratified sampling based on $K$-means clustering with all candidate variables~\cite{clustering_stratification}; 
    \item \textbf{SFS-KM}: Stratified sampling based on $K$-means clustering with variables selected by the conventional version of the sequential forward search~\cite{wrapper};
    \item \textbf{SFS-KM-V}: Stratified sampling based on $K$-means clustering with variables selected by our method (Algorithm~\ref{alg1}) for variance reduction. 
\end{itemize}
Here, the following sample allocation methods were implemented in stratified sampling: 
\begin{itemize}
    \item \textbf{Proportional}: Proportional sample allocation~\cite{stratifiedsampling} in Eq.~\eqref{eq:prop_alloc}; 
    \item \textbf{Optimal}: Optimal sample allocation~\cite{optimal} in Eq.~\eqref{eq:opt_alloc}.
\end{itemize}
Note that the sequential forward search~\cite{wrapper} in the SFS-KM method minimizes the within-cluster sum of squares of stratification variables, whereas our method (Algorithm~\ref{alg1}) in the SFS-KM-V method minimizes the variance of the sample mean of the outcome variable in Eq.~\eqref{eq:var}. 

We prepared training datasets for model estimation and testing datasets for performance evaluation. 
In the CUPED method, we chose the covariate that was most highly correlated with the outcome variable and then calculated its regression coefficient on the training dataset.
In the COSS method, we chose a covariate similarly to the CUPED method and then systematically extracted a sample of a specified size from a randomly drawn sample on the testing set.
In the stratified sampling methods, we performed clustering and sample allocation on the training dataset and then clustered the testing dataset for stratification based on the cluster centroids obtained from the training dataset.

We adopted the variance reduction rate as the evaluation metric in testing datasets. 
This metric indicates how much each method can reduce the outcome variance compared to the simple random sampling for testing datasets as 
\begin{equation}
\textbf{Variance reduction} \coloneqq \left(1-\frac{\mathrm{Var}(\hat{Y}_{\mathrm{red}})}{\mathrm{Var}(\hat{Y}_{\mathrm{rand}})}\right) \times 100,
\end{equation}
where $\hat{Y}_{\mathrm{red}}$ is the sample mean calculated by each variance reduction method, and $\hat{Y}_{\mathrm{rand}}$ is the sample mean calculated by the simple random sampling. 
The associated variances were estimated by repeating the calculation of the sample mean 10,000 times.

\subsection{Synthetic Datasets}

By following Hastie et al.~\cite{syntheticdata}, we generated synthetic datasets for the multiple linear regression model:
\begin{equation}
Y = \sum_{j \in [p]} \beta_j X_j + \varepsilon, 
\end{equation}
where $Y$ is an outcome variable, $X_j$ for $j \in [p]$ are stratification variables, and $\varepsilon$ is an error term. 
The ground-truth regression coefficients were defined by the following two patterns: 
\begin{itemize}
\item \textbf{beta-type 1}: $\beta_1=\beta_5=\beta_9=\beta_{13}=\beta_{17}=1$, and the other regression coefficients are $0$;
\item \textbf{beta-type 2}: $\beta_1=10$, $\beta_5=8$, $\beta_9=6$, $\beta_{13}=4$, $\beta_{17}=2$, and the other regression coefficients are $0$.
\end{itemize}

We also set the signal-to-noise ratio to 1.0, and the correlation parameter between explanatory variables to 0.35; see Hastie et al.~\cite{syntheticdata} for details on the dataset generation. 

\subsection{Results for Synthetic Datasets}

\begin{table}[t]
    \tabcolsep = 1mm
    \centering
    \caption{Variables selected in the synthetic datasets}
    \begin{tabular}{cll} \toprule
        Beta-type & Method & Variables \\ \midrule
        1 
        & CUPED & $X_9$\\
        & COSS & $X_9$\\
        & SFS-KM & $X_2$, $X_3$, $X_4$, $X_5$, $X_6$ \\
        & SFS-KM-V (Proportional, $n=10^4$) & $X_1$, $X_5$, $X_9$, $X_{13}$, $X_{17}$ \\ 
        & SFS-KM-V (Optimal, $n=10^4$) & $X_1$, $X_5$, $X_9$, $X_{13}$, $X_{17}$ \\ \midrule
        2
        & CUPED & $X_1$ \\
        & COSS & $X_1$ \\
        & SFS-KM & $X_2$, $X_3$, $X_4$, $X_5$, $X_6$ \\
        & SFS-KM-V (Proportional, $n=10^4$) & $X_1$, $X_2$, $X_5$, $X_9$, $X_{16}$ \\
        & SFS-KM-V (Optimal, $n=10^4$) & $X_1$, $X_2$, $X_5$, $X_9$, $X_{16}$ \\ \bottomrule
    \end{tabular}
    \label{tab:var_beta}
\end{table}

\begin{figure}[t]
    \centering
    \subfigure[$n=10^2$, beta-type 1]{\includegraphics[width=0.49\linewidth]{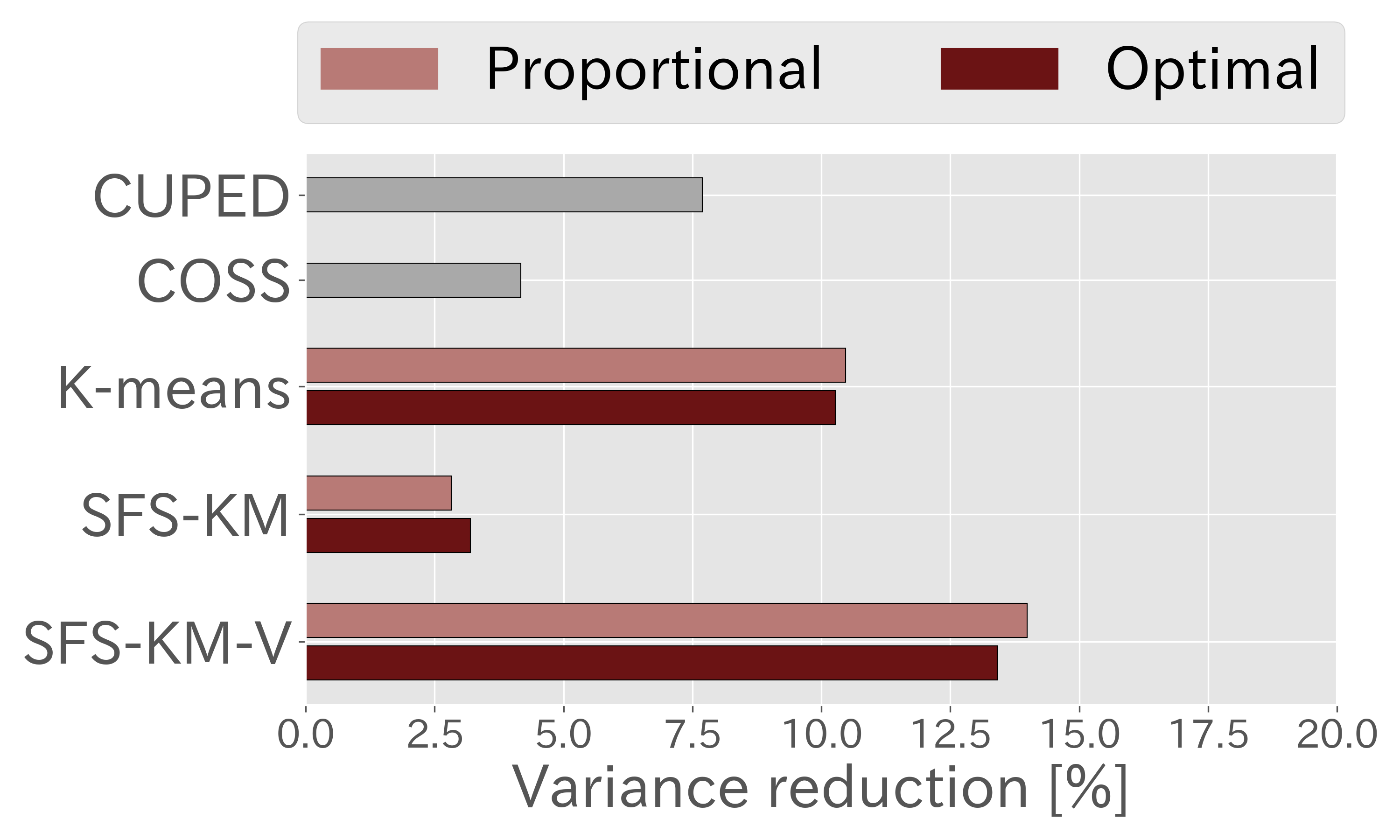}\label{fig:beta1_100}}
    \subfigure[$n=10^2$, beta-type 2]{\includegraphics[width=0.49\linewidth]{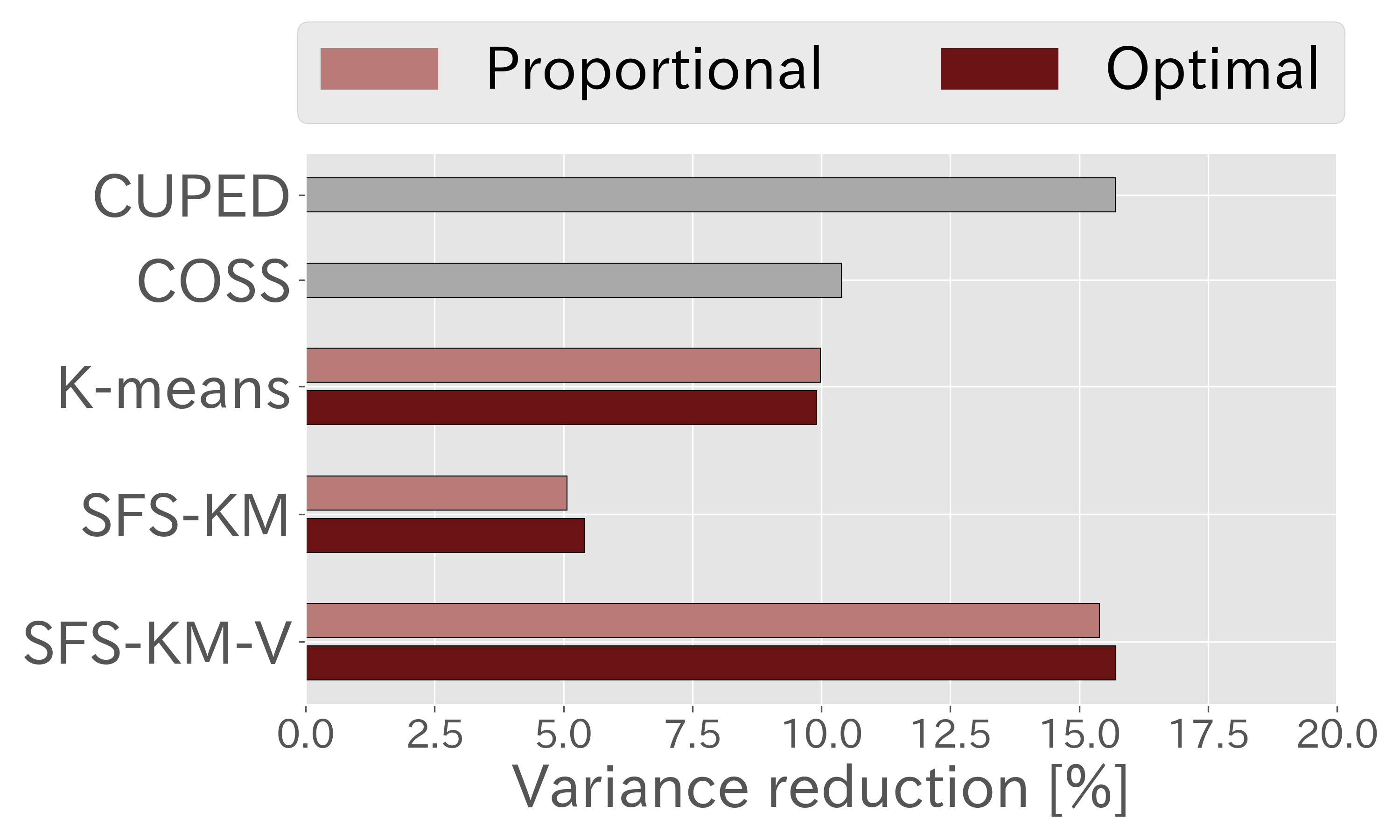}\label{fig:beta2_100}}
    \subfigure[$n=10^4$, beta-type 1]{\includegraphics[width=0.49\linewidth]{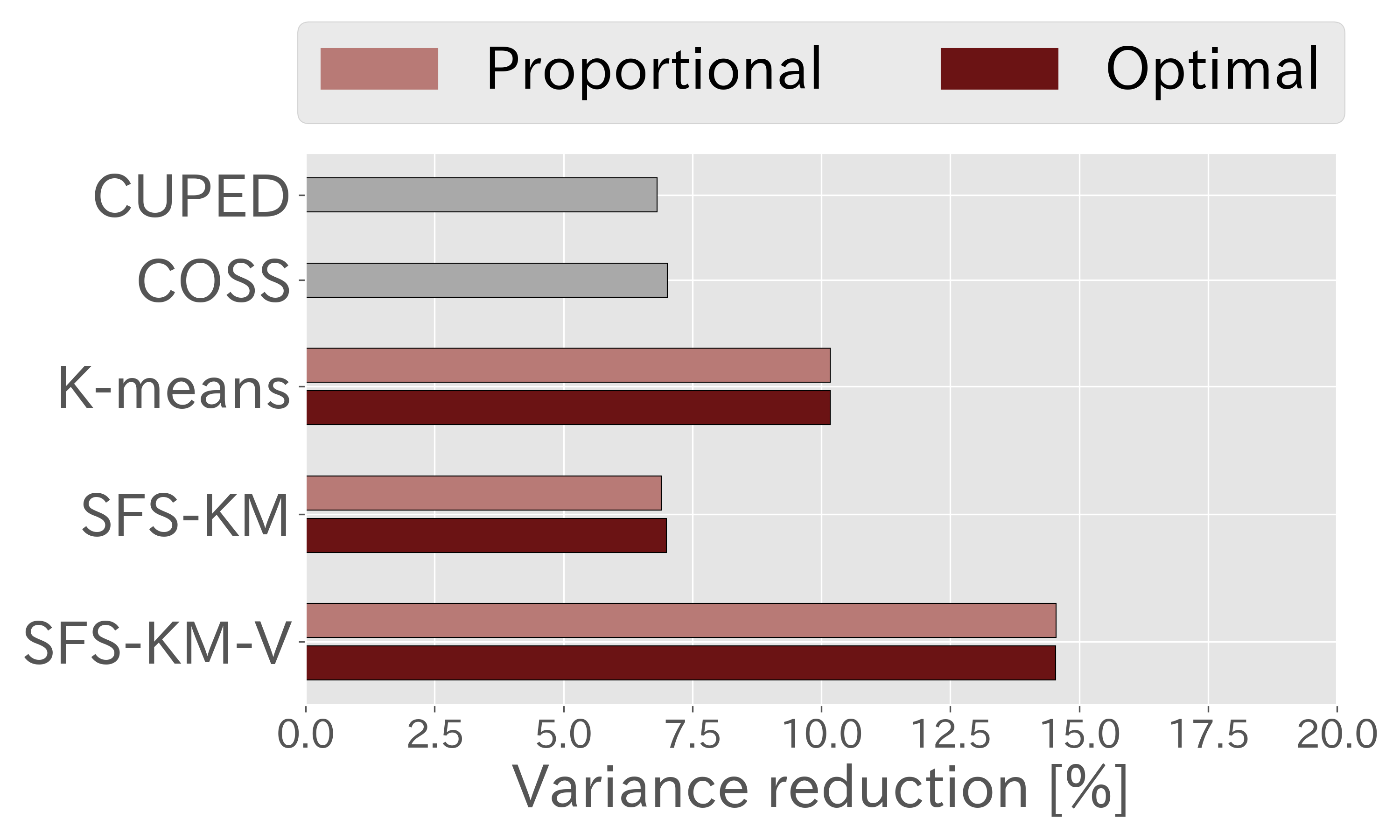}\label{fig:beta1_10000}}
    \subfigure[$n=10^4$, beta-type 2]{\includegraphics[width=0.49\linewidth]{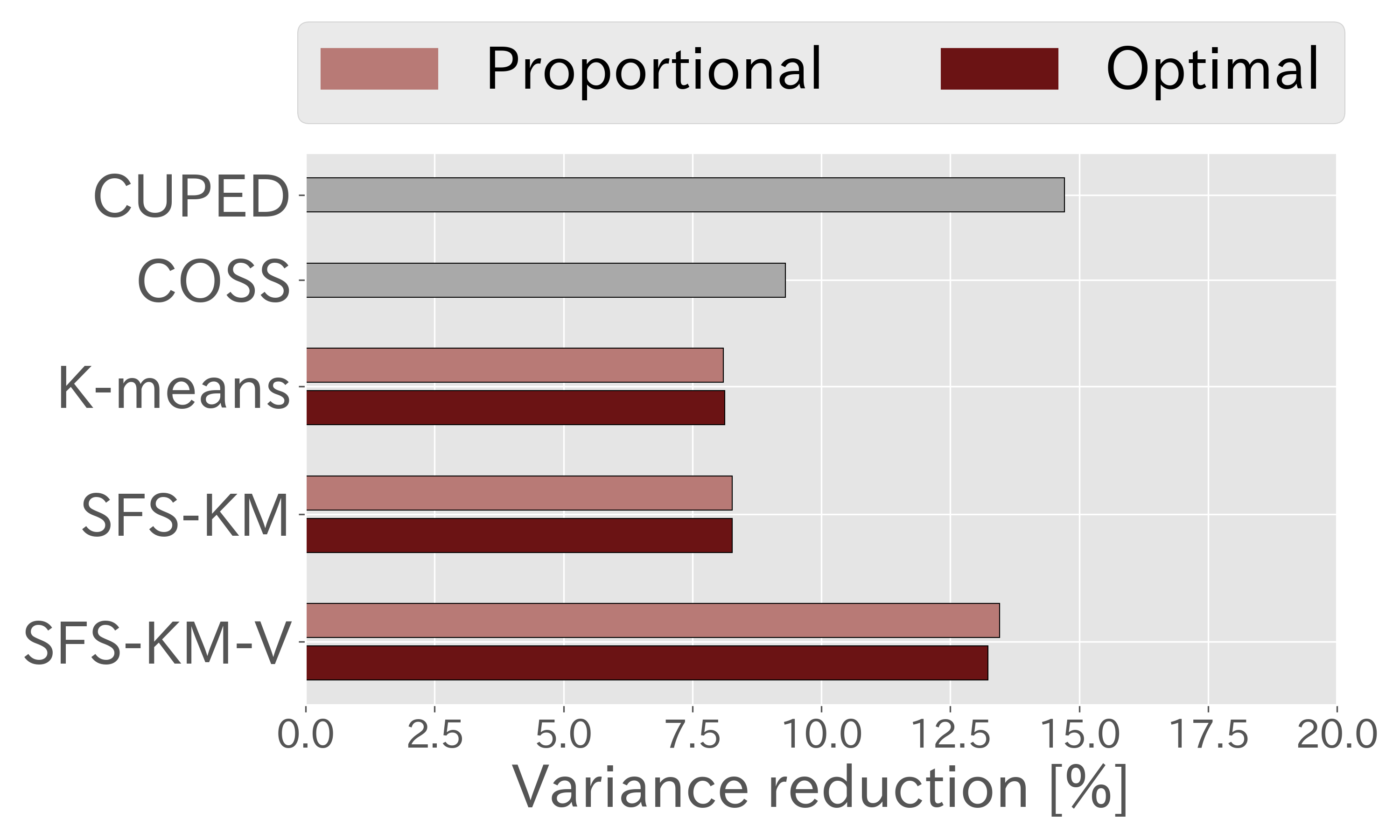}\label{fig:beta2_10000}}
    \caption{Variance reduction rates for the synthetic datasets}
    \label{fig:beta}
\end{figure}

\begin{figure}[t]
    \centering
    \subfigure[vs. $K$ ($\theta=5$, beta-type 1)]{\includegraphics[width=0.49\linewidth]{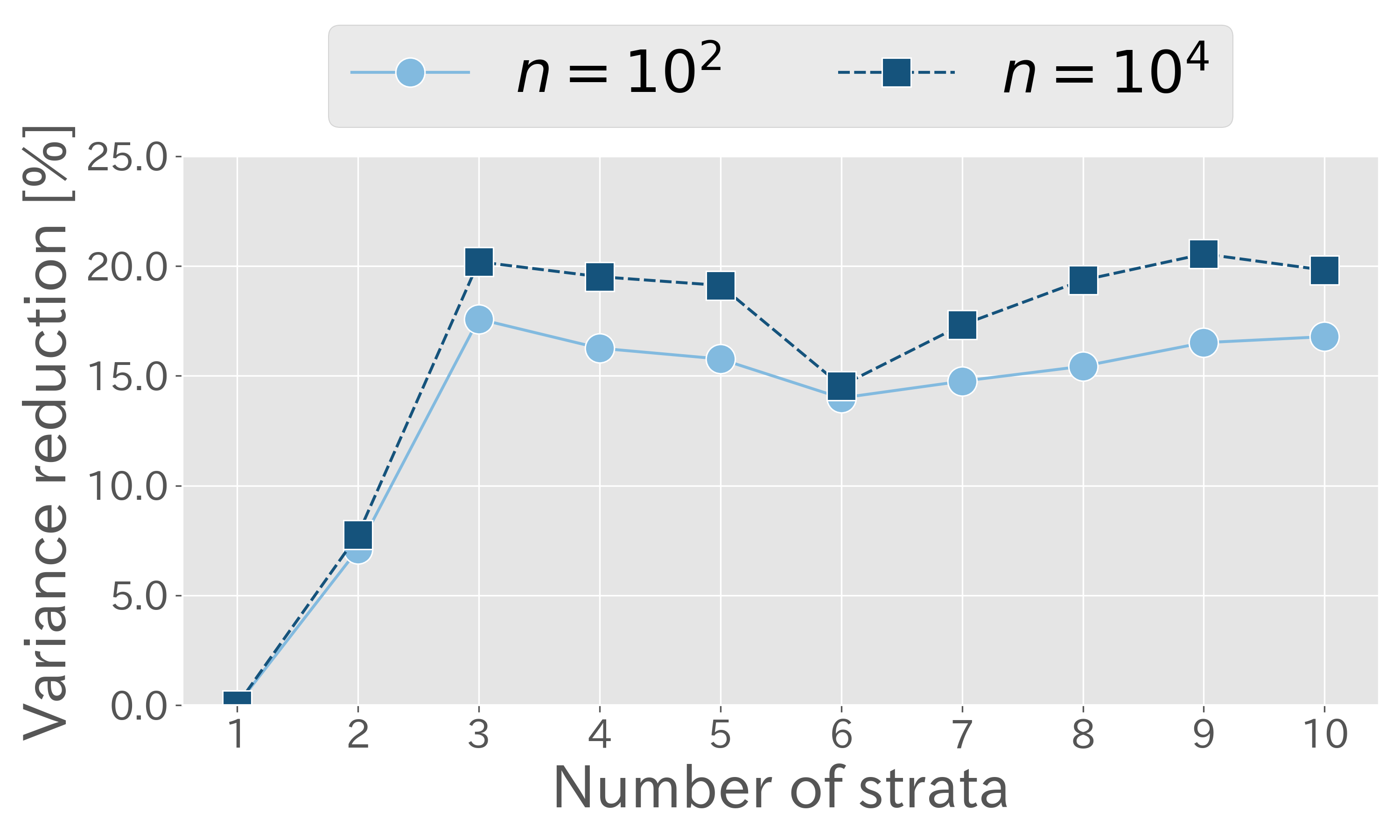}\label{fig:sensitivity_type1_K}}
    \subfigure[vs. $K$ ($\theta=5$, beta-type 2)]{\includegraphics[width=0.49\linewidth]{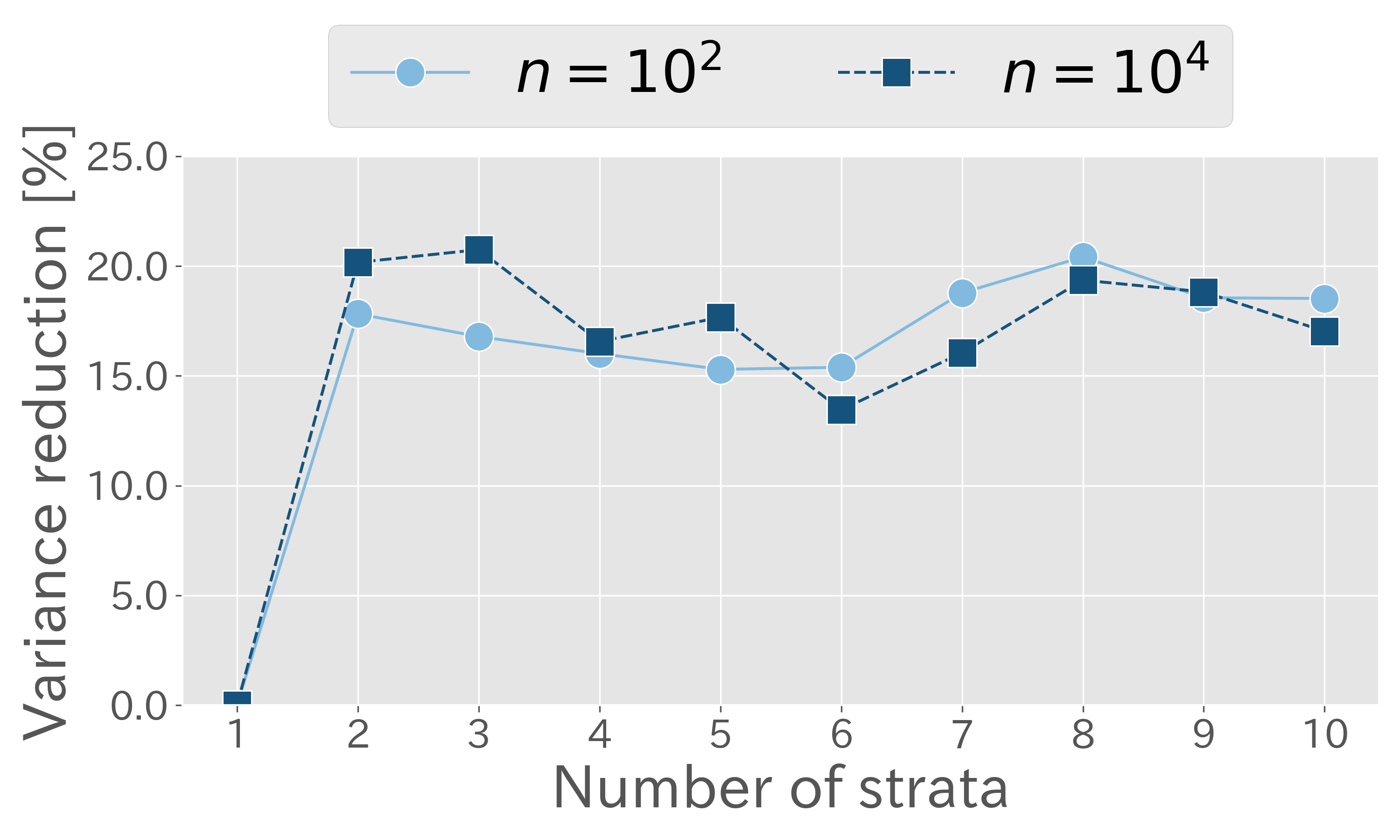}\label{fig:sensitivity_type2_K}}
    \subfigure[vs. $\theta$ ($K=6$, beta-type 1)]{\includegraphics[width=0.49\linewidth]{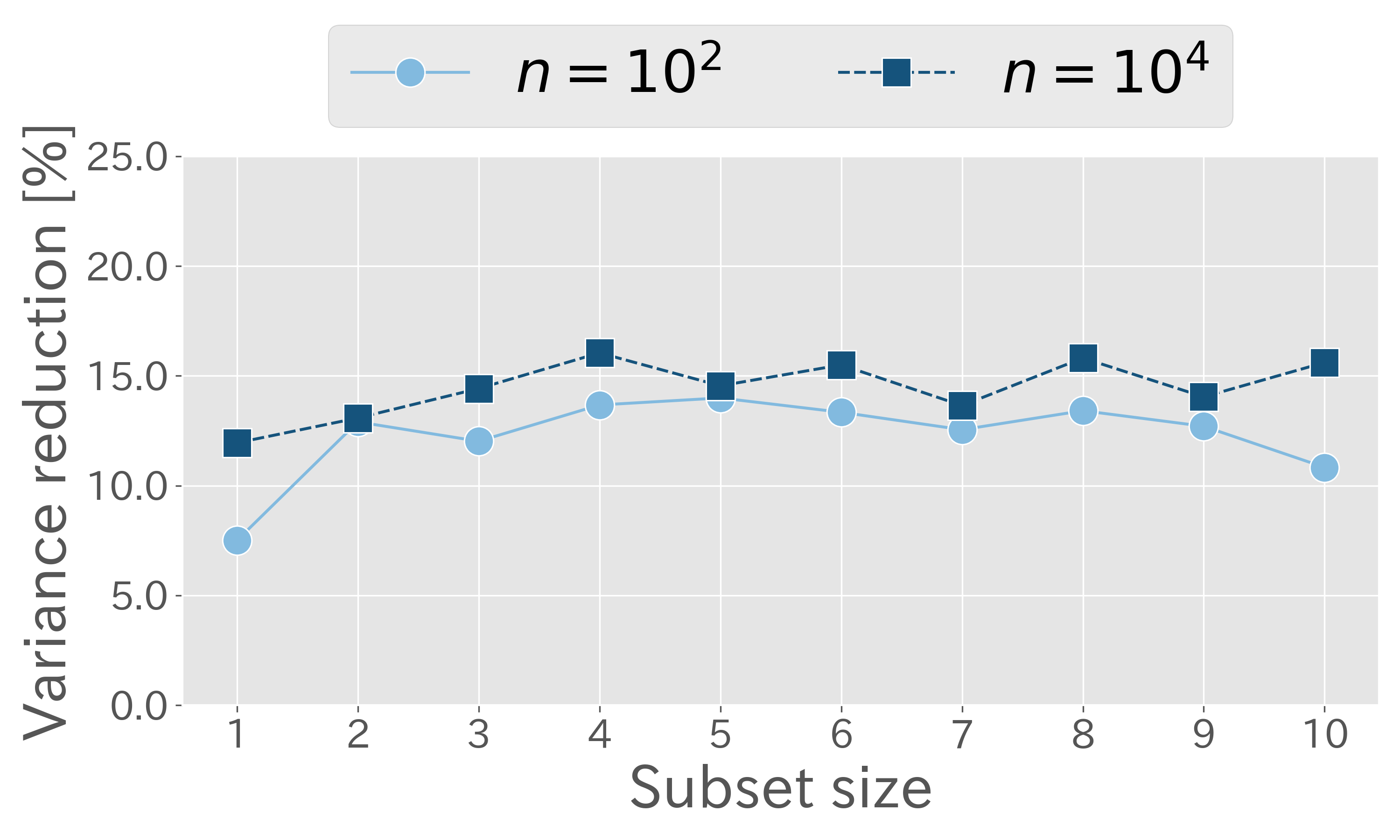}\label{fig:sensitivity_type1_theta}}
    \subfigure[vs. $\theta$ ($K=6$, beta-type 2)]{\includegraphics[width=0.49\linewidth]{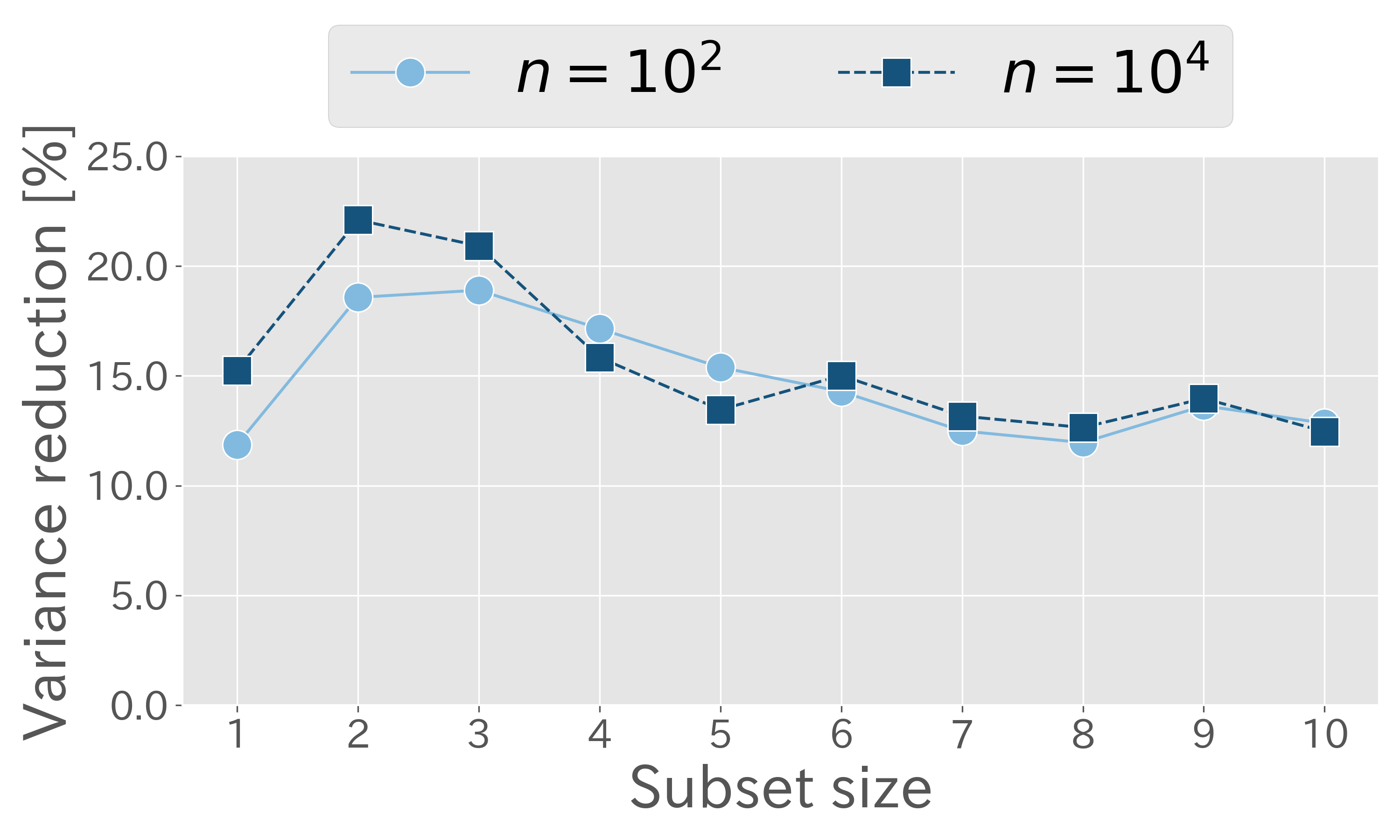}\label{fig:sensitivity_type2_theta}}
    \caption{Sensitivity analysis of variance reduction rates for the synthetic datasets}
    \label{fig:proportional}
\end{figure}

Fig.~\ref{fig:beta} shows the variance reduction rates of the five methods on the synthetic datasets with the sample size $n \in \{10^2, 10^4\}$, where the population size is $N=10^5$ for both training and testing, the number of strata is $K=6$, the number of candidate variables is $p=20$, and the subset size is $\theta = 5$.
Note that there was little difference between the proportional and optimal allocations in Fig.~\ref{fig:beta}, because the outcome variance $\sigma_k^2$ for each stratum $k \in [K]$ was equal in the synthetic datasets. 
Table~\ref{tab:var_beta} lists the variables selected in the synthetic datasets.

For the beta-type 1 pattern, our SFS-KM-V method achieved the highest variance reduction rates among all methods (Fig.~\ref{fig:beta}). 
In contrast, the SFS-KM method, which selects stratification variables without considering the outcome variable, performed poorly. 
Our SFS-KM-V method also selected the five variables with nonzero regression coefficients, whereas the SFS-KM method selected only one of the five variables with nonzero regression coefficients (Table~\ref{tab:var_beta}). 

For the beta-type 2 pattern, the performance of the CUPED and COSS methods was improved (Fig.~\ref{fig:beta}). 
In particular, the CUPED method performed as well as or slightly better than our SFS-KM-V method.
Although our SFS-KM-V method failed to select all the variables with nonzero regression coefficients, the CUPED and COSS methods selected the most influential variable (Table~\ref{tab:var_beta}). 

These results suggest that our method for stratified sampling is especially effective when there are multiple variables that have a certain correlation with the outcome variable as in the beta-type 1 pattern.
In contrast, the CUPED and COSS methods are relatively effective when there is only one variable that is highly correlated with the outcome variable as in the beta-type 2 pattern.

Fig.~\ref{fig:proportional} shows the variance reduction rate of our SFS-KM-V (Proportional) method as a function of $K$ (number of strata) and $\theta$ (subset size), with the same parameter configurations as in Figure 1.
No clear trend was observed regarding the effect of $K$. 
On the other hand, setting $\theta$ to a small value significantly improved the variance reduction rate for the beta-type 2 pattern.

\begin{figure}[t]
    \centering
    \subfigure[vs. $N$ ($p=20$)]{\includegraphics[width=0.49\linewidth]{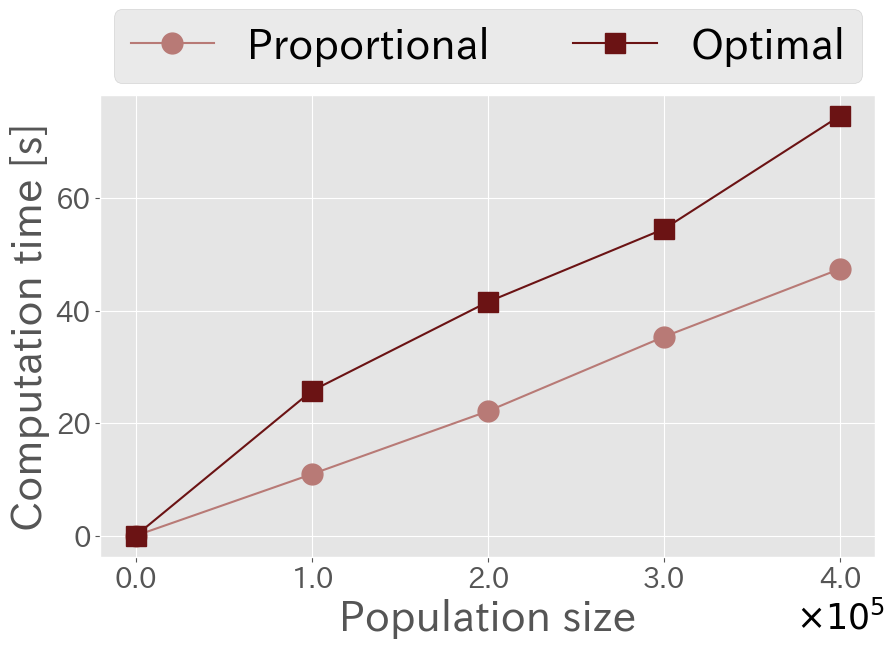}\label{fig:time_datasize}}
    \subfigure[vs. $p$ ($N=10^5$)]{\includegraphics[width=0.49\linewidth]{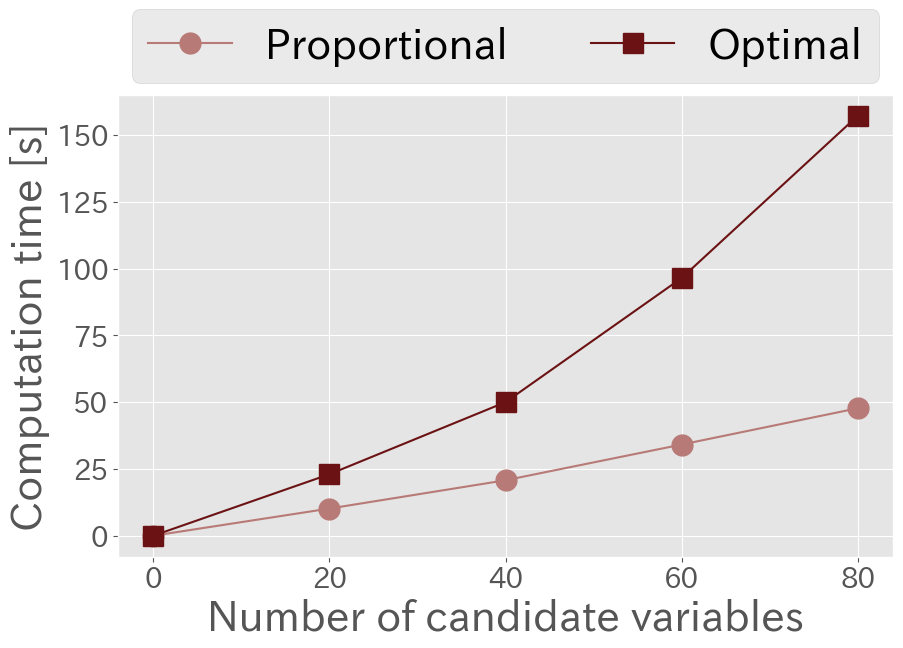}\label{fig:time_features}}
    \caption{Computation time for the synthetic datasets}
    \label{fig:time_beta}
\end{figure}

Fig.~\ref{fig:time_beta} shows the computation time required by our SFS-KM-V method as a function of $N$ (population size) and $p$ (number of candidate variables), where the sample size is $n=10^4$, the number of strata is $K=6$, and the subset size is $\theta = 5$.
Although the computation time was dependent on $p$ and $N$ (cf.~Section~\ref{sec:complexity}), our method can be executed in a reasonable time. 
For example, the computation time was about 25 s with the optimal sample allocation when $p=20$ and $N=10^5$. 

\subsection{Real-world Datasets}

We used the following two real-world datasets. 

\paragraph{GMV Dataset:}
We used actual data from a coupon distribution campaign that was conducted over four days on an online marketplace app operated by Mercari Inc., a Japanese e-commerce company. 
This dataset contains the gross merchandise volume (GMV) for each customer during the coupon validity period as the outcome variable, as well as 17 variables that represent each customer's purchase history prior to the coupon distribution (i.e., $p=17$).
The skewness of the outcome variable was 4.1, which indicates a highly skewed distribution.
We set $N=10^5$ for the population sizes for both training and testing.
We set $K=6$ for the number of strata and $\theta = 4$ for the subset size.

\paragraph{PM2.5 Dataset:}
We downloaded the PM2.5 Data of Five Chinese Cities, which contain hourly data in Beijing, Shanghai, Guangzhou, Chengdu, and Shenyang, from the UCI Machine Learning Repository\footnote{\url{https://archive.ics.uci.edu/}}.
We used seven quantitative variables (DEWP, TEMP, HUMI, PRES, Iws, precipitation, and Iprec), three qualitative variables (city, season, and cbwd), and one outcome variable (PM2.5 concentration). 
The skewness of the outcome variable was 2.8, which also indicates a highly skewed distribution.
Each qualitative variable was converted into dummy variables, resulting in a total of 21 variables (i.e., $p=21$).
The data observed in 2014 was used for training, and the data observed in 2015 was used for testing. 
After missing data removal, the population size was $N=43{,}800$ for both training and testing.
We set $K=5$ for the number of strata and $\theta = 5$ for the subset size.

\subsection{Results for Real-world Datasets}

\begin{figure}[t]
    \centering
    \subfigure[$n=10^2$, GMV]{\includegraphics[width=0.49\linewidth]{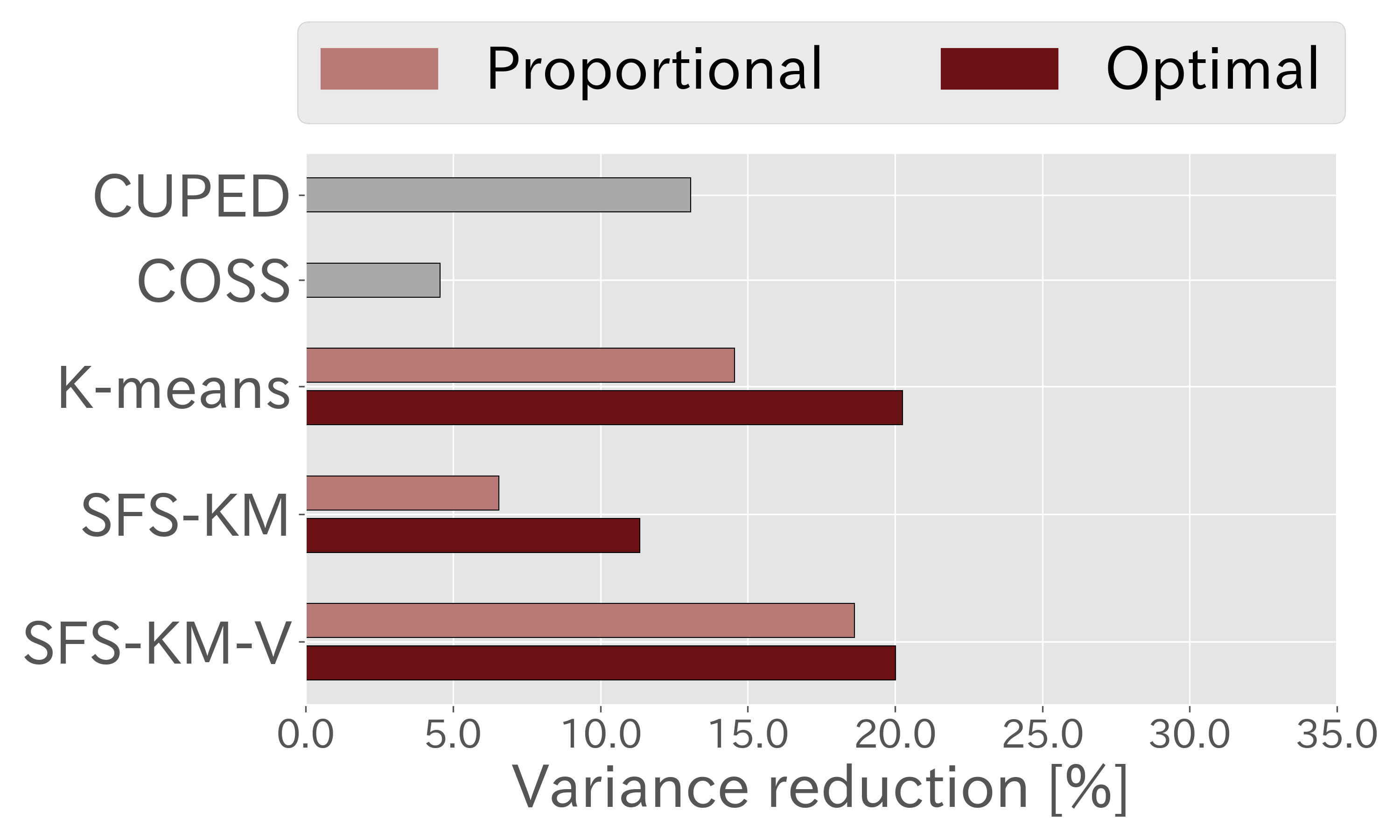}\label{fig:GMV100}}
    \subfigure[$n=10^2$, PM2.5]{\includegraphics[width=0.49\linewidth]{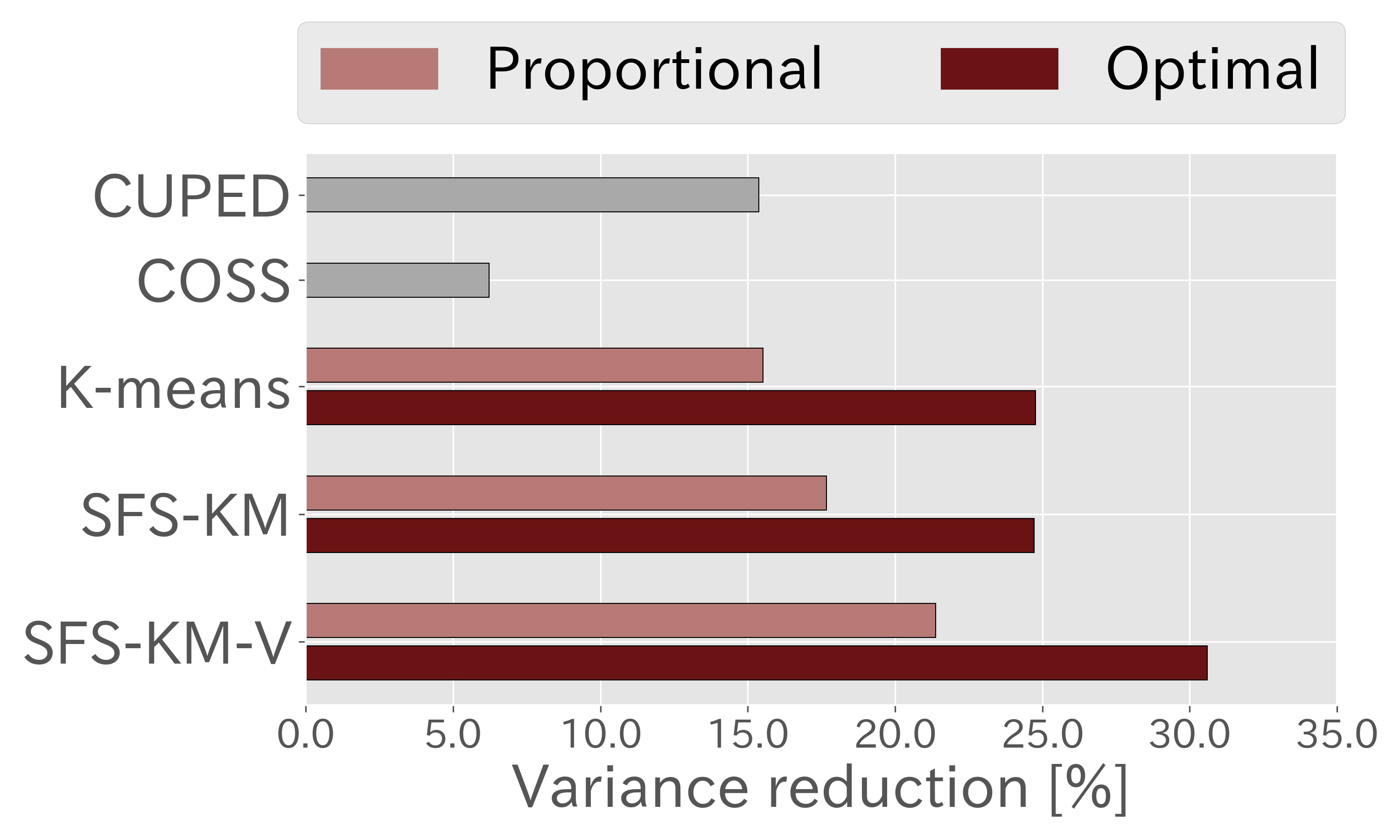}\label{fig:PM100}}
    \subfigure[$n=10^4$, GMV]{\includegraphics[width=0.49\linewidth]{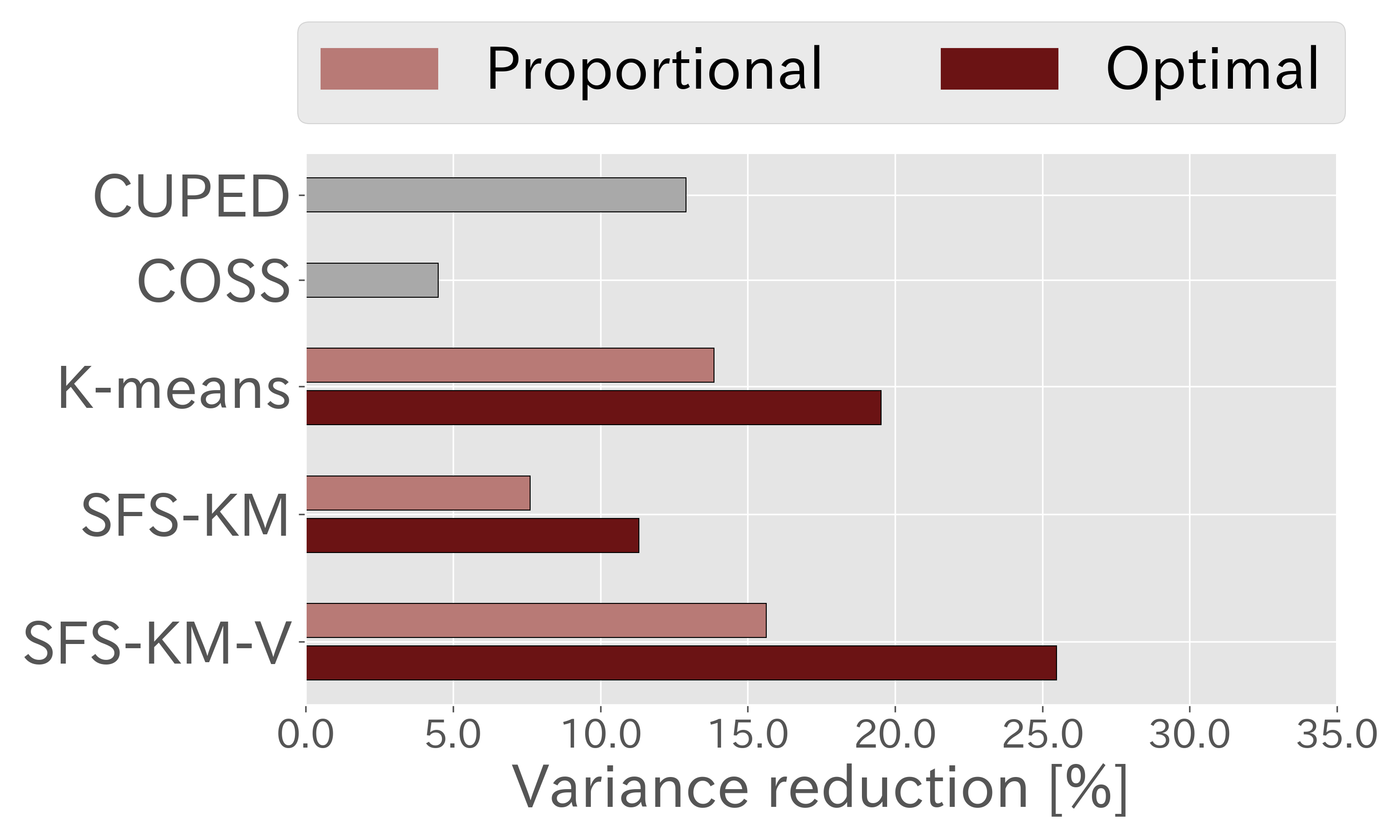}\label{fig:GMV10000}}
    \subfigure[$n=10^4$, PM2.5]{\includegraphics[width=0.49\linewidth]{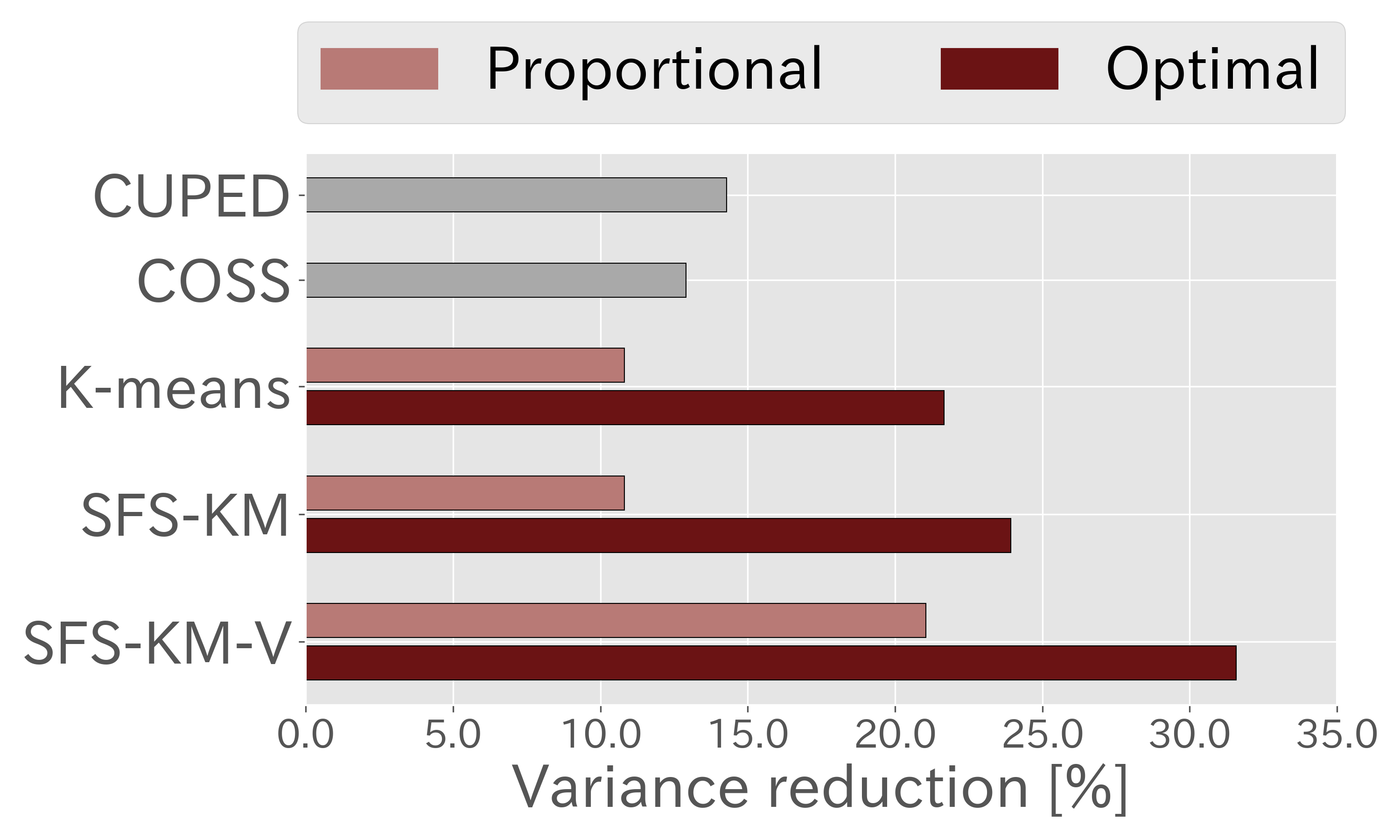}\label{fig:PM10000}}
    \caption{Variance reduction rates for the real-world datasets}
    \label{fig:real}
\end{figure}

Fig.~\ref{fig:real} shows the variance reduction rates of the five methods on the real-world datasets with the sample size $n \in \{10^2, 10^4\}$. 

First, we focus on the results of the K-means, SFS-KM, and SFS-KM-V methods for stratified sampling. 
Among these methods, our SFS-KM-V method achieved the highest variance reduction rates for each sample allocation for both sample sizes.
Additionally, the performance of stratified sampling was better with the optimal allocation than with the proportional allocation. 

Next, we compare the results of the SFS-KM-V method with the CUPED and COSS methods. 
Our SFS-KM-V method consistently achieved better variance reduction rates than did the CUPED and COSS methods. 
These results indicate the validity of our stratified sampling method, which can select a subset of variables suitable for stratified sampling in real-world datasets.

\section{Conclusion}

We proposed a computational framework to select an effective subset of variables used for stratified sampling in OCEs. 
Our algorithm selects stratification variables one by one by simulating a series of stratified sampling processes. 
We also estimated the computational complexity of our subset selection algorithm. 

We conducted computational experiments using synthetic and real-world datasets. 
In the experiments on the synthetic datasets, our method performed best when multiple variables were similarly correlated with the outcome variable, and also performed comparably to CUPED when a single variable was strongly correlated with the outcome variable.
In the experiments on the real-world datasets, our method clearly outperformed other methods in terms of the variance reduction rate.

A future direction of study will be to examine different types of subset selection techniques~\cite{featureselection} for clustering other than the sequential forward search~\cite{wrapper}.
Another direction of future research will be to incorporate clustering methods other than $K$-means clustering into our subset selection method. 
Stratification methods using decision trees were recently proposed~\cite{tabord2023stratification}, and we are considering comparison and integration of these tree-based methods with our method.
We are also planning to use our method to evaluate the impact of item ranking algorithms~\cite{uehara2024fast}. 

\begin{credits}
\subsubsection{\ackname} 
This work was partially supported by JSPS KAKENHI Grant Number JP25K01447. 

\subsubsection{\discintname}
The authors have no competing interests to declare that are relevant to the content of this article. 
\end{credits}
%
%
%
\bibliographystyle{splncs04}
\bibliography{cite.bib}
%

\end{document}